\documentclass[12pt]{article}
\usepackage{graphicx}

\begin{document}

\title{Three-body decay of \\ a rubidium Bose-Einstein condensate}
\author{J. S\"oding, D. Gu\'ery-Odelin, P. Desbiolles, F. Chevy, H. Inamori,\\
and J. Dalibard\\
Laboratoire Kastler Brossel\footnote{
Unit\'{e} de recherche de l'ENS et de l'Universit\'{e} 
Pierre et Marie Curie, associ\'ee au CNRS.}, Ecole Normale Sup\'{e}rieure,
\\
24 rue Lhomond, F-75231 Paris CEDEX 05, 
France} 

\date{\today}
\maketitle

\parskip 5mm

\begin{abstract}
We have measured the three-body decay of a 
Bose-Einstein condensate of rubidium
($^{87}$Rb) atoms prepared in the doubly polarized ground state $F=m_F=2$.
Our data are taken for a peak atomic density in the condensate
varying between $2\times 10^{14}$~cm$^{-3}$ at initial time and 
$7\times 10^{13}$~cm$^{-3}$, 16 seconds later. 
Taking into account the influence of the uncondensed
atoms onto the decay of the condensate, 
we deduce a rate constant for condensed atoms
$L=1.8\ (\pm\; 0.5) \times 10^{-29}$~cm$^{6}\;$s$^{-1}$. 
For these densities
we did not find a significant contribution of 
two-body processes such as
spin dipole relaxation.

\end{abstract}

\newpage

The remarkable achievement of Bose-Einstein condensation in alkali vapours
opens the way to many spectacular applications of ultra-cold atomic gases
\cite{Anderson95,Bradley97,Davis95}.
For most if not all of these applications, it is important to estimate the 
stability of the condensate with respect to inelastic processes, since the
condensate lifetime
ultimately determines the available time for a given experiment.

Two kinds of inelastic processes can contribute significantly to the decay
of a Bose -- Einstein condensate: three-body recombination and two-body spin relaxation.
In the first process, when three atoms are close to each other, two of them
may form a dimer or 
molecule, generally in an excited vibrational state, and the third atom
carries away the released energy. Since this energy is much 
larger than the typical
depth of the trap confining the atoms, the three atoms are lossed. 
For an atom in $\bf r$, the probability per unit time for this process 
is proportional to $n^2(\bf r)$, where $n(\bf r)$ is the spatial 
density of the atomic sample.
The second process can occur if the atoms are confined in a magnetic
trap.  In this case the atoms are 
prepared in a low-field seeking state, which is not the lowest
atomic state. The magnetic dipole interaction during a collision
between two trapped atoms
can lead to a spin flip which releases the Zeeman energy
$\Delta E \propto \mu_B B$, where $\mu_B$
 is the Bohr magneton and $B$ the local magnetic field. 
For this second process, the probability per unit time 
that an atom in $\bf r$
is expelled from the trap is proportional to $n(\bf r)$. 

In this paper we present experimental results concerning the decay
of a $^{87}$Rb condensate prepared in the doubly polarized ground state $F=m_F=2$
and confined in a magnetic trap.
We monitor the time evolution of the number of atoms in the condensate for
16 seconds, the condensed fraction remaining always higher than 40 \%.
We show that for our average densities in the condensate (between $4\times 
10^{13}$~cm$^{-3}$ and $12\times10^{13}$~cm$^{-3}$), the three-body 
recombination 
is the dominant loss process and we determine the corresponding rate 
coefficient. 
On the opposite we do not find evidence for two-body spin relaxation.

Our experimental setup is based on two glass cells, one positioned 
70 cm above the other.
Each is evacuated by a 25 l/s ion pump, and they are connected through 
a narrow glass tube ($\phi$ 9~mm, length 140 mm) to allow differential
 pumping. The lower cell is also connected to a titanium 
sublimation pump. This system allows us to produce a good
 vacuum in the lower cell (lifetime of the magnetic trap 
$\sim 40$~s) while having
sufficient Rb vapor pressure in the upper cell to load a 
magneto-optical trap 
(MOT) in $\sim 0.3$~s.

Light in the experiment is provided wholly by diode lasers 
at the rubidium resonance
(780~nm). The experimental sequence begins by loading $3\times 
10^7$ atoms into the upper MOT in 50~ms. These atoms are then 
pushed towards the lower cell by a 10~ms pulse from a vertical
 resonant laser beam. 
The measured final atomic velocity  is 14~m/s. The atoms are 
then recaptured in the lower MOT. The transfer efficiency is 
$70\;$\%, measured using a time of flight method based on 
the absorption of a probe beam located 2 cm below the center 
of the trap. The total duration of a loading cycle is 200~ms. 
We repeat this loading sequence 50 times while
 the lower MOT is operating.
After 10 seconds,  $10^9$ atoms are captured in the lower 
MOT\footnote{
By increasing the number of loading sequences to 100, we could 
load up to $1.6\; 10^9$ atoms, but this was not necessary for 
the experiments described here.}. 
The atoms are then cooled by a $10$~ms molasses phase and 
transferred into the magnetic trap.

This trap is of the Ioffe-Pritchard type, {\it i.e.} it is 
purely static and it consists of a nonzero local magnetic 
field minimum. The field is generated by three identical 
circular coils whose axes point towards $+x$, $-x$, $+y$
respectively ($z$ denotes the vertical axis), and whose 
centers are at equal distance 
from the center $O$ of the trap. The same current runs 
through the three coils.
These coils have a conical shape so that 
4 of the 6 lower MOT beams can be placed with a $\sim 45^\circ$ 
angle with respect to the horizontal plane (fig. 1). Each coil 
has 80 turns, is water cooled and can be run with 
a current of 100 A. For the experiments described below, a 
current of 46 A only was used.  
The leading terms in the magnetic field variations around $O$ 
are $(b'x,B_0+b''
(2y^2-x^2-z^2)/4,-b'z)$, with $B_0=12.9$~mT, 
$b'=1.39$~T$\,$m$^{-1}$, $b''=84.5$~T$\,$m$^{-2}$, for a 46~A 
current. These
quantities are accurately determined from the oscillation 
frequencies of the 
center of mass of the trapped atomic cloud.

At the end of the molasses phase, the atoms are optically 
pumped into the 
doubly polarized state $F=m_F=2$ (quantization axis $y$), 
and the magnetic trap is switched on. The transfer 
efficiency from the MOT to the magnetic trap is $50 \%$.
The atomic cloud is then further compressed by reducing 
the bias field $B_0$ to 
$0.126$~mT using a pair of Helmholtz coils aligned with 
the $y$-axis.
The transverse oscillation frequency $\omega_{x,z}\simeq 
(\mu_B b'^2/mB_0)^{1/2}$ increases to $2\pi \times 157$~Hz, 
while the longitudinal frequency $\omega_y=(\mu_B b''/m)^{1/2}$ 
remains equal to $2\pi \times 11.7$~Hz ($m$ is the atomic mass). 
At the end of this compression
stage, the temperature is 200$\;\mu$K.

We then perform forced evaporative cooling using a sweeped 
radio-frequency field
$\nu_{RF}$. The function $\nu_{RF}(t)$ is optimized so as to 
maximize 
the number of condensate atoms after a 18 second duration. 
The optimal
shape is found to be very close to an exponential law 
$\nu_a+\nu_b\; e^{-t/\tau}$
with a time constant $\tau=3.5$~s. The initial value 
for $\nu_{RF}$ is 15~MHz
and the final one is 0.900~MHz. This final value is 
10~kHz above the value 
$\nu_{\rm min}=0.890$~MHz which empties completely 
the trap. At this stage a condensate is formed.

The frequency $\nu_{RF}$ is subsequently kept constant 
for 2~s at 0.900~MHz 
to ensure that thermal equilibrium is reached and is then 
ramped up to 0.910~MHz in 
0.1~s. This corresponds to the initial time of the 
relaxation experiment. 
The atoms evolve now in the magnetic trap for an 
adjustable time $t$
between 0 and 16~s, with the RF on at the fixed value
0.910~MHz. This RF shield limits the trap depth to $\sim 0.8\ \mu$K, 
in order to ensure that
any atom which has been heated after an inelastic event leaves the trap
quasi-immediately.

The detection of the atomic cloud remaining in the trap 
after a relaxation time $t$
is made using an absorption imaging technique. 
The magnetic trap is switched off in $300\;\mu$s, and the 
atoms fall freely 
during $32$~ms. A $35\;\mu$s pulse from a weak 
($55\;\mu$W$\;$cm$^{-2}$) linearly 
polarized probe beam propagating along the $x$ axis then
illuminates the atomic cloud.
An optical system with a 1.64 magnification images 
the probe beam with the atomic cloud onto a CCD array. 
A second image with no atoms is taken 200~ms later to 
determine the laser intensity profile. The logarithm 
of the ratio of the two images yields
the cloud's optical density $d(y,z)=\sigma 
\int n_{\rm tot}({\bf r})\;dx$, where
$\sigma$ is the absorption cross-section and 
$n_{\rm tot}({\bf r})=n({\bf r})+
n_{\rm th}({\bf r})$ the total
(condensed +thermal) spatial density of the sample.

We choose a beam resonant with the atomic transition
$5s_{1/2}\rightarrow 5p_{3/2}$. 
To stay within the limits of the camera sensitivity, 
we make sure that the optical density remains below 3
(95~\% absorption). This limitation to relatively
dilute samples could be circumvented by taking an 
image with a laser beam detuned from the atomic resonance.
However, we found that the dispersive nature of the cloud 
for a non-zero detuning 
makes the accurate calibration of the images more delicate.

The value used for $\sigma$ takes into account the average 
intensity factor (7/15)
for the atomic transition, assuming that the atoms are 
uniformly spread over the 5 Zeeman sublevels of the 
$5s_{1/2}\ F=2$ ground level after the 32~ms free fall. 
When we analyze our data using the corresponding 
resonant cross-section value for $\sigma$, we find
a small ($10\%$) discrepancy between the 
measured BEC transition temperature $T_c^{\rm exp}$ 
with the predicted one $T_c^{\rm th}$ \cite{Giorgini96}:
 $T_c^{\rm exp}=
1.1\; T_c^{\rm th}$. 
We attribute it to a reduction of the absorption cross-section 
due to a finite linewidth of the probe laser, which 
results in an under-evaluation of the number of atoms.
The number of atoms given in the following have been 
corrected to take into
account this imperfection of the imaging laser.

A typical image consists in two features: (i) A central elliptic
region of high density corresponds to the condensate cloud. At $t=0$,
the total size of this region is
$\Delta y \times \Delta z= 130\;\mu$m$\times 270\;\mu$m. 
(ii) A slightly 
larger quasi-isotropic region corresponds to the uncondensed fraction
of the atomic cloud.
The analysis of the images is made by a 3-step fitting procedure. 
In a first step
we fit the total image with the sum of a Gaussian function, 
representing the
uncondensed fraction, and a function corresponding to the
 integration along the $x$ axis of a paraboloidal distribution. 
The latter describes in the Thomas-Fermi limit
\cite{Dalfovo98}
 the equilibrium density profile of the condensate within 
the harmonic trap, and it is known to remain valid (with 
a scaling factor) after a
ballistic expansion \cite{Castin96}. The quality of this 
first fit is subsequently improved in two steps where (i)
 we fit only the part of the image which contains no condensate
again by a Gaussian
function to derive the temperature $T$, (ii) we substract 
the Gaussian distribution
derived in this way and we fit only the central component 
to derive precisely the number of condensed atoms $N$. The
 total number of atoms $N_{\rm tot}$ 
is finally evaluated
from the integral of $d(y,z)$ over the whole 
image\footnote{
We have also developped a more sophisticated procedure 
for the determination of the temperature 
assuming an ideal Bose-Einstein distribution for the
 uncondensed fraction rather
than a Maxwell-Boltzmann distribution. Since an accurate
  determination of the temperature ({\it i.e.} to better
 than 10\%) is not essential for the present work,
we have used here only the simpler fitting procedure.}.

To accurately measure the decrease of the number $N$ of atoms in the
condensate, drifts and fluctuations in the initial number of condensed 
atoms and in the temperature have to be suppressed as far as possible. 
The most important limitation to experimental accuracy is the drift of 
the magnetic field minimum $B_0$. 
A decrease of $B_0$ by only $10^{-6}$~T will diminish 
$\nu_{\rm min}$ by 7\,kHz. Because the cold atomic samples are prepared 
by evaporating down to a {\em fixed} final rf frequency, this decrease 
corresponds to 
a temperature increase of $\sim (h/k_B)\,$7\,kHz$\,=\,330$\,nK. This
value has to be compared to our typical transition temperature 
$T_c \approx 200$\,nK. In order to keep the initial
fraction of condensed atoms
stable around 75\%, the magnetic field minimum $B_0$ has 
to be stable to within less than a few $10^{-7}$~T. This is a very stringent
condition because the value $B_0\approx\,10^{-4}$~T is obtained by
subtracting two large ($\sim\,10^{-2}$~T) nearly equal fields.

In normal operation we have observed drifts in the value of $B_0$ 
on the order of $10^{-5}$~T. We have identified temperature changes of 
the magnetic
field coils and their support structure as primary source for these 
drifts. Temperature changes lead to thermal dilations 
by a few $\mu$m that can account for the observed drifts. 
To get rid of these drifts we have automatized the experiment and
data acquisition completely. This allows us to leave the laboratory while 
the computer is producing one condensate every $\sim$35\,s, varying the
relaxation time of the cloud in the magnetic trap automatically.
After 15 minutes the setup is in thermal equilibrium and 
drifts of $B_0$ are below $3 \times 10^{-7}$~T 
during one data series ($\sim$1.5 hours). 

To reduce systematic errors due to residual drifts in $B_0$ even further, 
we have put a set of 16 different relaxation times $t$ in random order and   
repeated this same set 9 times in a row, thus preparing $16\times 9$
condensates in total. For each relaxation time we measure the number of
atoms remaining in the condensate. Finally, we calculate their
average $\bar{N}(t) = (1/9) \sum N_i(t)$.

The number $\bar{N}(t)$ of condensate atoms is shown in fig.\,2 
on a logarithmic scale. 
The marked non-exponential decay indicates that in the beginning the atoms 
are lost mainly by two- or three-body inelastic collisions.
The initial and final numbers of atoms in the condensate are
respectively 300\,000 and 17\,000, and the
initial and final average condensate 
densities are $11.8\times 10^{13}$~cm$^{-3}$ and $3.8\times
 10^{13}$~cm$^{-3}$. The temperature of the sample
is constant at $75\;(\pm 13)$~nK over the measured time interval.

We now compare the measured atom number 
in the condensate with the solution of 
the differential equation governing the time 
evolution of $N$.
In a first step we consider only collisions 
between condensate atoms.
The loss rates due to two- and three-body collisions
 are described by terms 
$ -G \,\int n^2({\bf r}) \;d^3 r = -G \langle n \rangle N$ 
and  
$ -L \,\int n^3({\bf r}) \;d^3 r = -L \langle n^2 \rangle N$, 
respectively, where
$G$ and $L$ are the rate coefficients for two- and three-body 
collisions, and where
we put for a function $\eta({\bf r})$: 
$\langle \eta\rangle=\int \eta({\bf r})\; n({\bf r})\; d^3r\;/\;N$.
The integrals can be calculated in the Thomas-Fermi
limit of a parabolic condensate density, giving
$\langle n \rangle=c_2 N^{2/5}$ and $\langle n^2 \rangle=c_3 N^{4/5}$,
with $c_2=(15^{2/5}/(14\pi)) \left(m \bar{\omega}/
(\hbar \sqrt{a})\right)^{6/5}$
and $c_3=(7/6)c_2^2$. Here $a=5.8\;$nm denotes the 
scattering length 
and $\bar\omega=(\omega_x\omega_y\omega_z)^{1/3}$. 
The validity of the Thomas-Fermi
approximation can be checked in fig.~3, where we have plotted the
full width $\Delta z$ of the condensate  as a function of $N$, together
with a fit using the modelling function $\Delta z=\Delta z_1\; N^{1/5}$. 
We find $\Delta z_1=21.2\ \mu$m, 
in very good agreement (to within $4\%$) with
the Thomas-Fermi prediction $\Delta z_1=20.4\ \mu$m  \cite{Castin96}.

We have not been able to find an analytical 
solution to the differential equation describing both two- {\em and} 
three-body collisions at the same time, but we can solve the equation 
in the presence of either two- or three-body collisions:
\begin{equation}
\frac{1}{N}\frac{dN}{dt} =- G c_2 N^{2/5} -\frac{1}{\tau} , 
\end{equation}
\begin{equation}
\frac{1}{N}\frac{dN}{dt} = - L c_3 N^{4/5} -\frac{1}{\tau}\,\,.
\end{equation}

The dotted and dashed lines in fig.\,2 show solutions of the 
differential equations for two- and three-body collisions that 
fit the experimental data best. We find 
$
\{
\tau=\infty\;,\; 
G=2.78\ (\pm 0.02)\; \times 10^{-15}\;{\rm cm}^3{\rm s}^{-1}
\}$
for two-body collisions and
$\{
\tau=14.8\,{\rm s}\; , \; L=2.23\ (\pm 0.11)\; \times 10^{-29}\;
{\rm cm}^6{\rm s}^{-1}
\}$
for three-body collisions, where the error is 
purely statistical at this stage. 
The lifetime of the condensate $\tau$ is determined by small-angle
collisions with the thermal background gas and expected to be smaller
than the lifetime of an uncondensed cloud. Indeed 
a small energy on the order of $100\,$nK 
(the chemical potential) is sufficient to eject the atoms out of
the condensate, whereas an energy of $\sim 1\,$mK is necessary
to make an atom leave the magnetic trap. We found a 
lifetime for thermal clouds $\tau_{\rm th}\sim 40$\,s, 
which contradicts the two-body hypothesis. 

The error bars in the inset show the statistical 
error for $\bar{N}$. The noise on the number of atoms 
can be read directly from the length of the error bars and it amounts to
only 1\%. The difference between the experimental data and the best fit
is shown with dashed and dotted lines.
Whereas the residue of the fit to a three-body 
decay law (dashed line) stays most of the time within the bounds
given by the statistical error, the fit to a two-body decay law
has deviations by several $\sigma$ for many points. On the
basis of our model, a pure two-body decay (including $-1/\tau$)
can therefore be ruled out with certainty, whereas a pure 
three-body decay is well compatible with the data.

More quantitatively, the expected mean value of a $\chi^2$ 
distribution with 13 degrees of freedom (16 data points minus 
3 fit parameters) is 13. The $\chi^2$ value for the two-body fit 
is $\chi^2=235$, corresponding to a rejection confidence
 much below $10^{-2}$, whereas the $\chi^2$ value for three-body decay 
of $\chi^2=12.7$ corresponds to 
what is to be expected statistically.

Our analysis so far relies on solving the differential equation 
for $N(t)$, which implies two drawbacks:\ \  
(1)~It neglects collisions of condensate atoms with the 
thermal component of the cloud. This is not a very good approximation, 
because the central density of the thermal cloud 
$n_{\rm th}({\bf r}={\bf 0})$
reaches $17\%$ of the mean condensate density 
$\langle n \rangle$ in the measured time interval.\ \ 
(2)~We can not take account of both two-body and three-body 
collisions at the same time, making it impossible to place 
a stringent upper limit on the two-body rate coefficient $G$.

In presence of collisions with the thermal cloud with density
$n_{\rm th}({\bf r})$ the differential equation for $N(t)$ 
has to be modified:
\begin{equation}
\frac{1}{N}\frac{dN}{dt} =
\;-\; L \left[  \langle n^2\rangle 
	  + 6\, \langle n \;n_{\rm th} \rangle 
	  + 6\,  \langle n_{\rm th}^2 \rangle 		\right]
\;-\; G \left[ \langle n\rangle + 2\, \langle n_{\rm th}\rangle  \right] \
\;-\; \frac{1}{\tau}\;.
\label{dN/Ndt}
\end{equation}
The first term on the right hand side represents three-body
collisions, with three, two or one condensate atom participating in the 
collision (and beeing subsequently lossed). 
Similarly, the second term on the right hand
side represents two-body collisions with two or one condensate
atoms participating.
In writing this equation we have taken into account the 
role of the symmetry of the condensate wave function with respect to
particle exchange
\cite{Kagan85,Burt97}. 

In view of the excellent quality of the data we now calculate 
the time derivative of $\mbox{ln}\,N$ (equal to $dN/Ndt$) 
and its statistical error directly from the experimental data 
\footnote{
For every set of three
adjacent times $t_{i-1}, t_i, t_{i+1}$ we find the parabola that 
passes through the three data  points ln$\bar{N}_{i-1}$, ln$\bar{N}_i$,
ln$\bar{N}_{i+1}$ and take the slope of the parabola at time $t_i$
as the value of $d $ln$N/dt$ at time $t_i$.}. The result is
plotted in fig.\,4 as a function of $f(N,T)$, the term 
multiplying $L$ in \ eq.\,(\ref{dN/Ndt}): 
\begin{equation}
f(N,T) =   \langle n^2\rangle 
	  + 6\, \langle n\; n_{\rm th} \rangle 
	  + 6 \, \langle n_{\rm th}^2 \rangle 	
\quad,
\label{f}
\end{equation}
which is evaluated in the appendix.
Points to the right correspond to large atom numbers and short 
relaxation times.
If two-body collisions are negligible, eq.\,(\ref{dN/Ndt})
tells us that $dN/Ndt$ depends linearly on $f(N,T)$. 
Fig.\,4 reveals exactly this linear dependence, the fit yielding 
$\tau=18.0\;(\pm 1.0)$\,s and 
\begin{equation}
L=1.81\ ( \pm 0.06\  \pm 0.40)\times 10^{-29}\ {\rm cm}^6\,{\rm s}^{-1}
\end{equation}
The first error represents the {\em statistical} error, 
while the second, dominating error gives the {\em systematic} 
error of 20\% due essentially to the calibration of the number 
of atoms detected.

We can now place an upper limit on the two-body rate coefficient
by fitting $dN/Ndt$ as a function of $f(N,T)$
and $g(N,T)=\langle n\rangle + 2\langle  n_{\rm th}\rangle $.
The fit yields $L=2.7\;( \pm 0.5)\times 10^{-29}$cm$^6$s$^{-1}$ and
$G=-1.8\;( \pm 1.0)\times 10^{-15}$cm$^3$s$^{-1}$. Hence we conclude that
the two-body rate $G$ is smaller than $10^{-15}$cm$^3$s$^{-1}$ for 
$B=0.13$~mT.

In the previous analysis we have assumed that for each three-body
collision, only the three atoms which are directly involved in the
collision are lost from the trap. In fact, secondary collisions 
between condensate atoms and fast atoms emerging from a three-body
collison might augment the loss of condensate atoms per three-body 
collision to a value larger than three. We calculate the mean 
free path of a fast atom by estimating the collisional cross section
from a semiclassical model, 
and obtain $\sigma <10^{-12}$~cm$^2$ for a relative
velocity larger than 10~ms$^{-1}$. 
The mean free path ($>80\;\mu$m) is then larger than the transverse size of 
the condensate ($8\;\mu$m full width), 
so
that this effect can not change 
the results of our analysis.

We now briefly compare our results with previous measurements
and predictions. The three-body recombination rate has been 
measured for Rb atoms prepared in the low hyperfine state $F=-m_F=1$ 
\cite{Burt97}. The result for that state is 
$L=5.8\times 10^{-30}$~cm$^6$s$^{-1}$ for condensed atoms, {\it i.e.} 3
times smaller than the result found here for the stretched state $F=m_F=2$. 
Two theoretical predictions have been made for the three-body recombination
rate of Rb, based on different assumptions. A first calculation, based on 
the Jastrow approximation, led to a very small value 
$L=0.7\times 10^{-30}$~cm$^6$s$^{-1}$
for condensed atoms \cite{Moerdijk96}. A second prediction considers the case
of a large and positive scattering length $a$ for the binary elastic collision, 
corresponding to the existence of a weakly bound state
\cite{Fedichev96}. The recombination
rate to this bound state is then shown to be  $L=3.9\, \hbar a^4/(2m)=
1.6\times 10^{-30}$~cm$^6$s$^{-1}$, with $a=5.8$~nm \footnote{
We deduce this number from \cite{Fedichev96} by taking 
into account the reduction by a factor 6 for condensate
atoms, and the fact that the three atoms are lossed 
in an 3-body recombination event.}.
Note that the hypothesis at the basis of \cite{Fedichev96} is not strictly
valid for Rb atoms. Clearly more theoretical work 
is needed to give a quantitative
account of the measured rate, for both hyperfine states. 
Concerning the two-body spin relaxation it has been pointed 
out \cite{Julienne97}
 that it should be
anomalously small ($1\; -\; 2\times 10^{-15}$cm$^{3}$s$^{-1}$ 
for a field of 
$10^{-4}$~T) due to the 
coincidence of the scattering lengths for elastic collisions in the 
states $F=m_F=2$ and $F=-m_F=1$; our result confirms this prediction.

In conclusion, we have presented measurements of the inelastic collision
rate of a magnetically trapped Bose-Einstein 
condensate of Rb atoms in their upper hyperfine state. 
Our analysis, which includes the influence of the
thermal component of the atomic cloud, allows us to determine 
the value of the rate coefficient of three-body collisions 
and to put a low upper limit on the rate coefficient 
for inelastic two-body collisions. We have discussed possible
limitations of our analysis by additional loss mechanisms and
find them to be negligible.

J.\,S. acknowledges support by the Alexander von Humboldt-foundation.
This work was partially supported by CNRS, Coll\`{e}ge de France,
DRET, DRED and EC (TMR network ERB FMRX-CT96-0002).

\newpage
\section*{Appendix:
Decay rate from a condensate in presence of a thermal component}
\vskip 1cm

The decay rate of the condensate via a $p$-body 
collision may involve either
a collision between $p$ condensate atoms, or a 
collision between $q$ ($<p$) condensate
atoms and $p-q$ atoms from the thermal fraction. Taking into account 
symmetrization
\cite{Kagan85}, this decay rate can be written:
\begin{equation}
{\dot N}=-G\left( \int n^2({\bf r})\; 
d^3 r+2\int n({\bf r})\;n_{\rm th}({\bf r})\; d^3r\right)
\label{twobodyapp}
\end{equation}
for a two-body process and 
\begin{equation}
{\dot N}=-L\left( \int n^3({\bf r})\; d^3 r+6
\int n^2({\bf r})\;n_{\rm th}({\bf r})\; d^3r + 6\int n({\bf r})
\;n^2_{\rm th}({\bf r})\; d^3r \right)
\label{threebodyapp}
\end{equation}
for a three-body process.
As indicated in the text we calculate the first integral of 
each of these two expressions
in the Thomas-Fermi limit, assuming a parabolic condensate density
$$
n({\bf r})= (\mu-V({\bf r}))/g  
$$
inside the condensate, and $n({\bf r})=0$ outside. Here $\mu$ is the chemical potential, $V({\bf r})$ denotes
the harmonic trapping potential and $g=4\pi \hbar^2 a/m$.

To evaluate $n_{\rm th}({\bf r})$ in 
the remaining terms, we use the Hartee-Fock 
approximation \cite{Dalfovo98}.
Since the density of the thermal fraction is 
always smaller than the central density
of the condensate by an order of magnitude, we 
neglect the effect of the thermal component on the condensate distribution. 
In this approximation the density of the thermal 
fraction is given by:
$$
n_{\rm th}({\bf r})=\Lambda_T^{-3}\; g_{3/2}
\left(e^{-|\mu-V({\bf r})|/k_BT} \right)
$$
where $\Lambda_T=h/\sqrt{2\pi m k_B T}$ and 
$g_{3/2}(z)=\sum z^\ell \ell^{-3/2}$. This 
expression takes into account the repulsion 
of the uncondensed atoms from
 the condensate by the interaction potential 
$2gn({\bf r})$. The overlap integrals entering 
in (\ref{twobodyapp}) and 
(\ref{threebodyapp}) are then calculated numerically.

The final results can be cast in the form:
$$
\frac{\dot N}{N}= -L\left(\langle n^2\rangle +
6 \,\langle n\rangle \;\tilde n_{\rm th}({\bf 0}) \; \alpha(\tilde \mu)
+ 6\, \tilde n^2_{\rm th}({\bf 0}) \; \beta(\tilde \mu)\right)
\ -G \left(\langle n\rangle + 2 \,\tilde n_{\rm th}({\bf 0})
\gamma(\tilde \mu)
  \right)
$$
with $\tilde \mu=\mu/(k_B T)$.
The quantity $\tilde n_{\rm th}({\bf 0})=
\Lambda_T^{-3}g_{3/2}(1)$ represents the 
uncondensed density of an ideal Bose gas at
 temperature $T$ (below $T_c$)
at the center of the trap. The functions 
$\alpha, \beta,\gamma$ are equal to 1 in the
 limit $\mu \ll k_B T$ and smaller than 1 
otherwise, since the overlap between 
$n({\bf r})$ and $n_{\rm th}({\bf r})$ is then reduced. 
For instance, for 
$\mu=k_B T$ (experimental situation at $t=0$), we
 find $\alpha(1)=0.26$,
$\beta(1)=0.11$ and $\gamma(1)=0.31$. The effect 
of mixed collisions (condensate +
thermal fraction) is therefore notably reduced with 
respect to the ideal gas case.

\newpage 

\begin{figure}
\begin{center}
\includegraphics[height=4in]{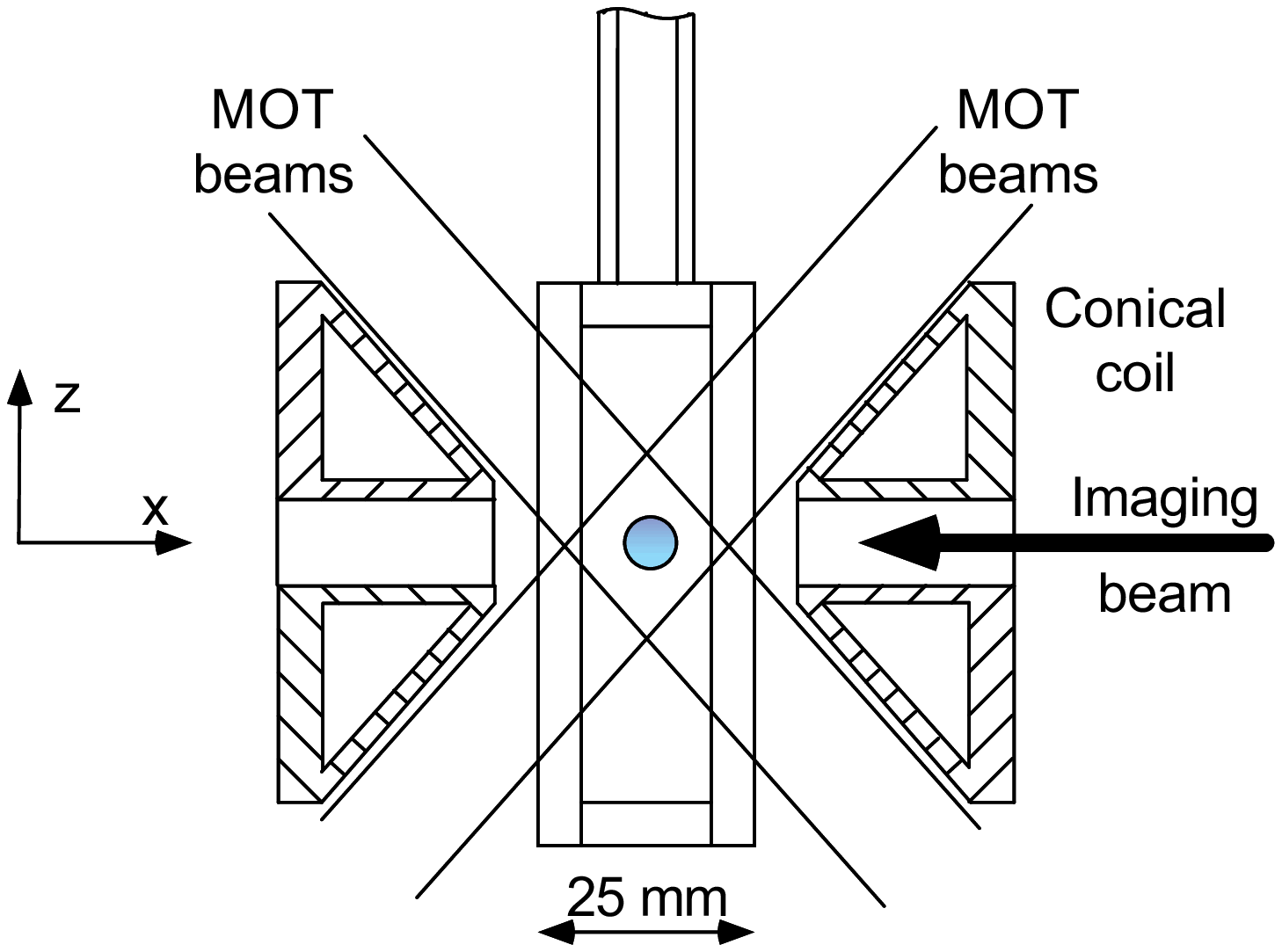}
\end{center}
\caption[]{Cut through the lower glass cell and two of the three
conical magnetic field coils generating the trapping potential. 
The third pair
of MOT beams is perpendicular to the plane of the figure.} 
\label{fig1} 
\end{figure}

\newpage

\begin{figure}
\includegraphics[height=4in]{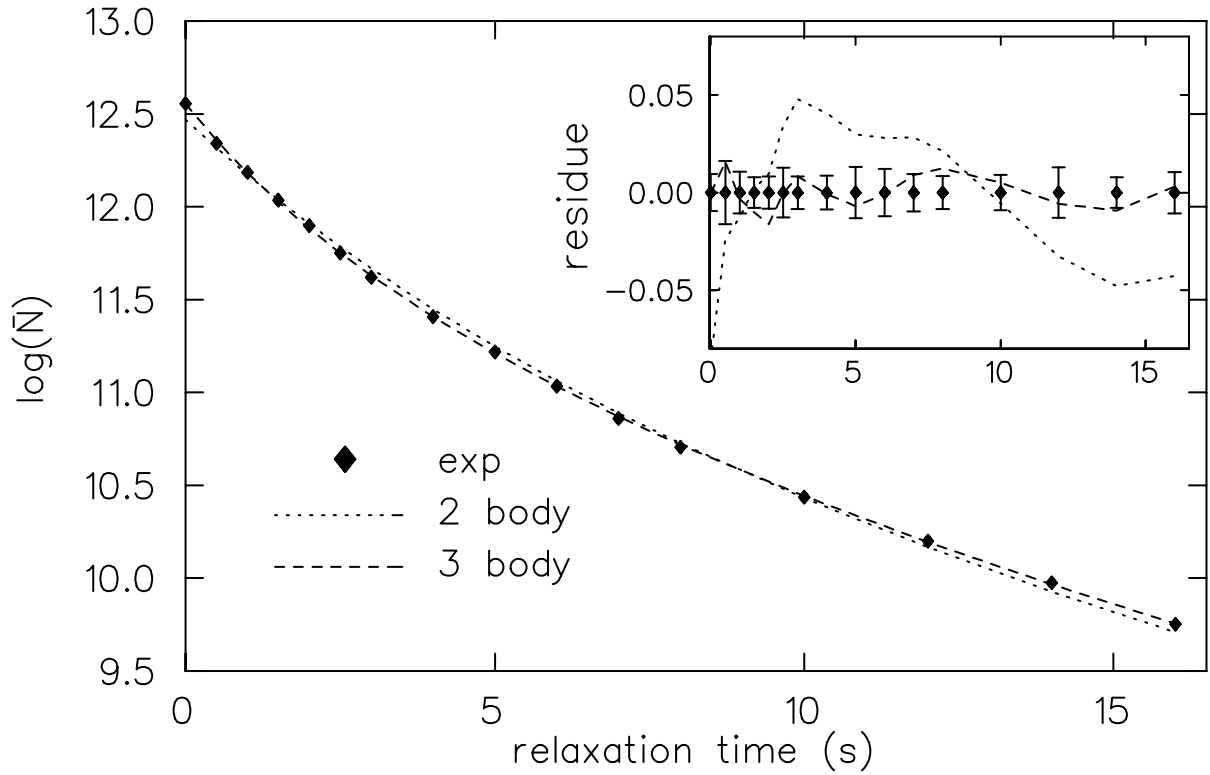}
\caption[]{Logarithm of total number of atoms remaining in the 
condensate after a relaxation time $t$. Each point is
an average over 9 measurements. The inset shows the deviation of the 
experimental points from a fit to a two-body decay law (dotted line) and
to a three-body decay law (dashed line). The error bars indicate the
statistical error for each point.}
\label{fig2} 
\end{figure}

\newpage

\begin{figure}
\begin{center}
\includegraphics[height=5in]{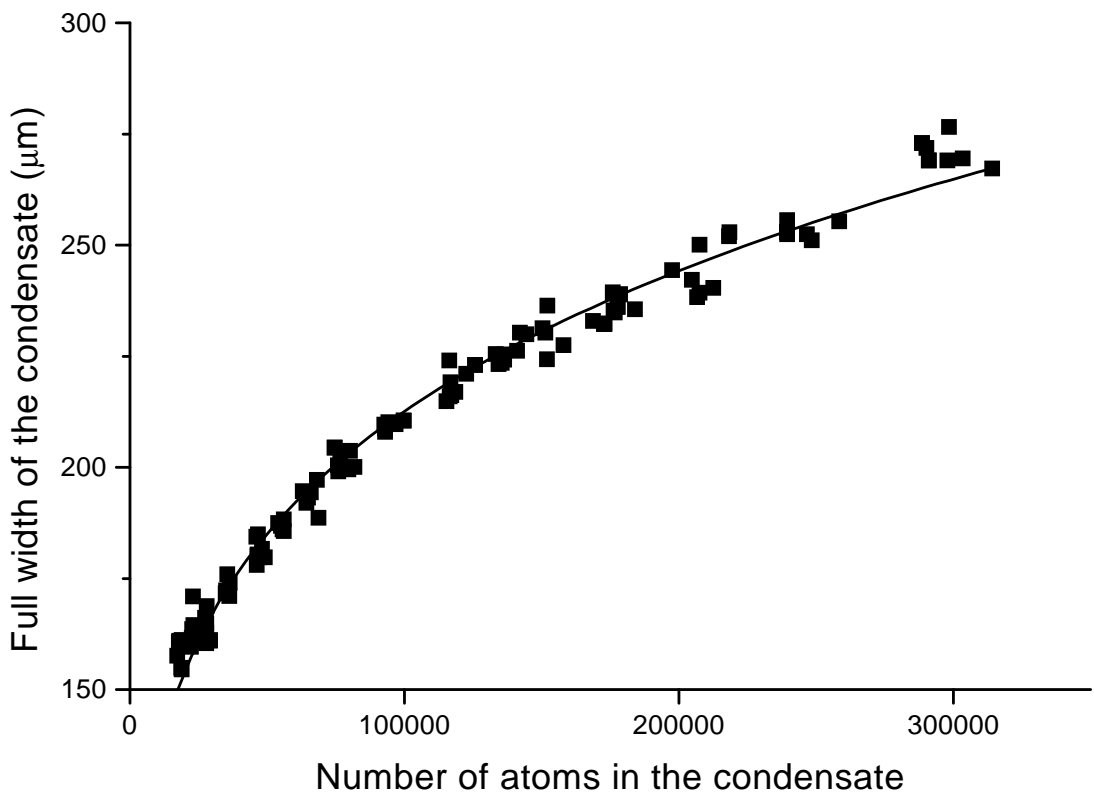}
\end{center}
\caption[]{Variations of the full width $\Delta z$ of the 
condensate along the $z$ axis, as a function of the number 
of condensed atoms $N$. The full line
is a fit with the function $\Delta z=\Delta z_1\, N^{1/5}$.}
\label{fig3} 
\end{figure}

\newpage

\begin{figure}
\begin{center}
\includegraphics[height=3.5in]{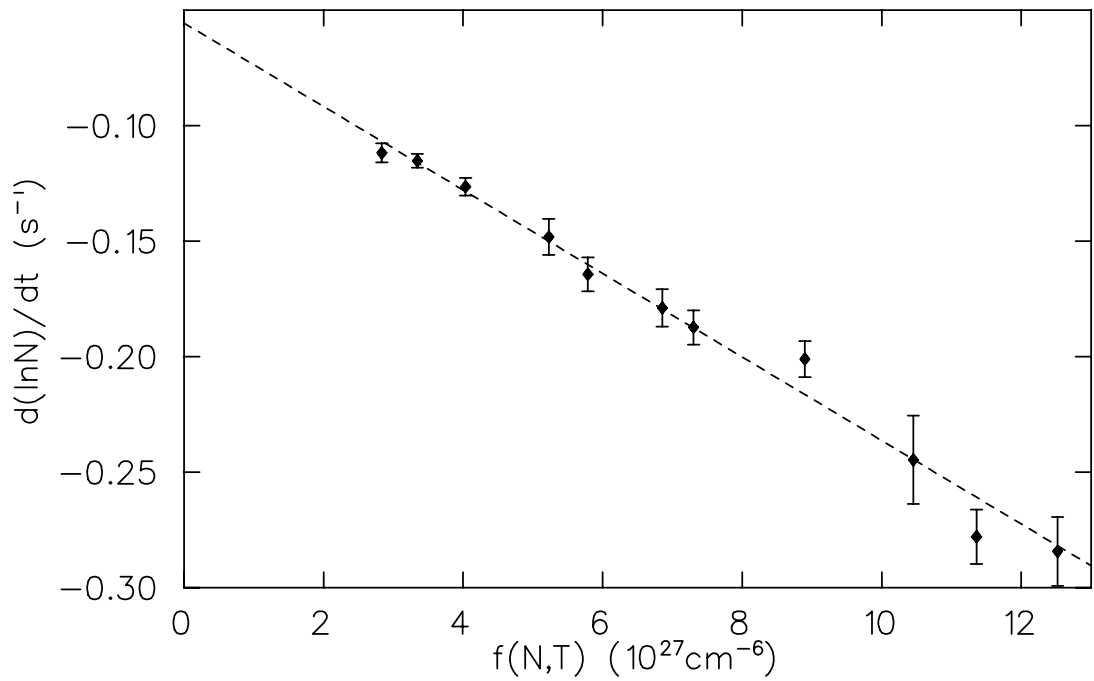}
\end{center}
\caption[]{Logarithmic derivative of the measured number of 
condensate atoms as a function of $f(N,T)$ (see eq.\,\ref{f}).
The function $f$ is chosen to give a linear dependence for a 
three-body decay even in the presence of a nonnegligable thermal
background. The dashed line shows this linear fit.}
\label{fig4} 
\end{figure}

\end{document}